\documentclass[12pts]{article}
\usepackage{CJK}
\usepackage{fancyhdr}
\usepackage{bm}
\usepackage{amsmath}
\usepackage{color}
\usepackage{graphicx}
\usepackage{subcaption}
\usepackage{multirow}
\usepackage{algorithm}
\usepackage{algorithmic} 
\usepackage{authblk}

\usepackage{graphicx}
\usepackage{setspace} 
\newcounter{ass}

\usepackage{array} 
\title{A mixture logistic model for panel data with a Markov structure}
\author[$\dagger$]{Yu-Hsiang Cheng}
\author[$\star$]{Tzee-Ming Huang\thanks{Corresponding Author: tmhuang@nccu.edu.tw}}
\affil[$\dagger$]{Department of Finance, Shih Hsin University, Taiwan (R.O.C.)}
\affil[$\star$]{Department of Statistics, National Chengchi Univeristy, Taiwan (R.O.C.)}
\begin{document}
\maketitle
\begin{abstract}
In this study, we propose a mixture logistic regression model with a Markov structure, and consider the estimation of model parameters using maximum likelihood estimation. We also provide a forward type variable selection algorithm to choose the important explanatory variables to reduce the number of parameters in the proposed model. 
\end{abstract}

\section{Introduction}
In this study, we propose a mixture logistic regression model with a Markov structure, and consider the estimation of model parameters using maximum likelihood estimation. To reduce the number of parameters in the model, we also propose a variable selection algorithm. Some results of this study, including the model framework and the E-M algorithm for computing the maximum likelihood estimators of model parameters, have been included in the report of a MOST project\cite{Cheng:2022}, which was written in Chinese.  The variable selection algorithm is given in this article and not in \cite{Cheng:2022}.
The proposed model can be used to analyze panel data sets containing observations of a discrete response variable $Y$ and some explanatory variables $X_1$, $\ldots$. $X_p$ at time points $1$, $\ldots$, $T$ for $n$ individuals. It is assumed that $Y$ can have $K$ states: $1$, $\ldots$, $K$. There are two features of the proposed model. One is that the model can be applied when the $n$ individuals are from $L$ groups and the group labels are not observed. The other feature is that within each group, each transition probability that the response variable $Y$ moves from  current state to the next state is allowed to depend on $X_1$, $\ldots$, $X_p$ and the current state of $Y$.

To describe the proposed model, we first introduce some notation. Let $\mathbf{X} = (X_1, \ldots, X_p)$. For $i=1$, $\ldots$, $n$, let $(Y_t^{(i)}, \mathbf{X}_t^{(i)})$ denote the observed value  of $(Y, \mathbf{X})$ at time point $t$ for the $i$-th individual for $t \in \{ 1, \ldots, T\}$, and let $D^{(i)}$ denote the group membership of the $i$-th individual.  Then the proposed model is described as follows.
For $t = 2, \ldots, T$, $i = 1, \ldots, n$,  $u =1, \ldots, K$, $v = 2 , \ldots, K$, and $\ell = 1$, $\ldots$, $L$,
\begin{equation} \label{mixture1} 
 \log\Bigg(\frac{P(Y_t^{(i)} = v|Y_{t-1}^{(i)} = u, \mathbf{X}_t^{(i)},  D^{(i)} = \ell)}{P(Y_t^{(i)} = 1|Y_{t-1}^{(i)} = u, \mathbf{X}_t^{(i)}, D^{(i)} = \ell)}\Bigg) 
=\bm{\alpha}_{u, v, \ell}^\top \mathbf{X}_t^{(i)}
\end{equation}
where $\bm{\alpha}_{u, v, \ell}$ is a column vector of $p$ constants depending on $u$, $v$, $\ell$ only. 
 
Based on past studies on logistic regression for panel data, one may include the time effect using a random effect model or a fixed effect model (see Ten Have et al. \cite{TenHave:2000} for example). However, by analyzing some real data sets, we found that the observed response for an individual was affected by past response values (of the same individual), but not affected by the time point of observation. Therefore, we consider the Markov structure in (\ref{mixture1}) to describe the time effect. 

 The rest parts of this article are organized as follows. In Section \ref{sec:em}, we give the details of parameter estimation procedure of the proposed model, and describe how to obtain a clustering result based on the proposed model. In Section \ref{sec:var_sel}, we propose a variable selection algorithm to reduce the number of parameters in the model and present a simulation result to demonstrate the performance of the variable selection algorithm. 

\section{Parameter estimation via E-M algorithm and clustering} \label{sec:em}
For parameter estimation of the proposed model in (\ref{mixture1}), we obtain the maximum likelihood estimators of the model parameters using the E-M algorithm, which is a common approach for finding maximum likelihood estimators of parameters in mixture models. A general description of the E-M algorithm can be found in Dempster et al.\cite{Dempster:1997}.  For the E-M algorithm, we take the observed data $\bm{Y} = (Y_1^{(1)}, \ldots$, $Y_T^{(1)}$, $\ldots,  Y_1^{(n)}, \ldots Y_T^{(n)})$, 
 $\bm{X} = (\bm{X}_1^{(1)}, \ldots, \bm{X}_1^{(n)}, \ldots, \bm{X}_T^{(1)}, \ldots \bm{X}_T^{(n)})$ and the unobserved data $\bm{Z} = (D^{(1)}$, $\ldots$, $D^{(n)})$ as the complete data. To write down the complete data likelihood, for $v \in \{ 1, 2, \ldots, K \}$, vectors $\mathbf{x}$, $\bm{\alpha}_2, \ldots, \bm{\alpha}_K$ in $R^p$, define
\[
 P_{v}(\mathbf{x}; \bm{\alpha}_2, \ldots, \bm{\alpha}_K) = \left\{
   \begin{array}{ll}
   \frac{\exp\big(\bm{\alpha}_v^\top \mathbf{x}\big)}{1+\sum_{s=2}^K \exp\big(\bm{\alpha}_s^\top \mathbf{x}\big)}, & \mbox{ if } v \neq 1;\\
   \frac{1}{1+\sum_{s=2}^K \exp\big(\bm{\alpha}_s^\top \mathbf{x}\big)}, & \mbox{ if }  v = 1.
  \end{array} \right.
\]
For $\ell = 1, \ldots, L$, let $\pi_{\ell} = P(D^{(i)} = \ell)$
and
\[
 A_{i, \ell} = \prod_{t=2}^T \sum_{u=1}^K \sum_{v=1}^K P_{v}\Big(\mathbf{X}_t^{(i)}, \bm{\alpha}_{u, 2, \ell}, \ldots, \bm{\alpha}_{u, K, \ell}\Big)\mathbf{I}\Big(Y_{t-1}^{(i)} = u\Big)\mathbf{I}\Big(Y_t^{(i)} = v\Big)
\]
for $i=1$, $\ldots$, $n$. 
Let $\bm{\alpha}^*_\ell = \{\bm{\alpha}_{u, v, \ell}|u, v \in \{1, \ldots, K\}\}$, 
$\bm{\theta} = \{\pi_\ell, \bm{\alpha}^*_\ell|\ell = 1, \ldots, L\}$, then the logarithm of 
the complete likelihood function is given by
\[
 \mathcal{L}(\bm{\theta}|\bm{Y}, \bm{X}, \bm{Z}) = \sum_{i=1}^n\sum_{\ell=1}^L \mbox{I}(D^{(i)}=\ell)\Big(\log(\pi_\ell)+\log(A_{i, \ell})\Big).
 \]
Then, the parameter estimators can be obtained by carrying out the E-step and the M-step iteratively in the E-M algorithm. Suppose that $\bm{\theta}^{(s-1)}$ is the estimated parameter vector after the $(s-1)$-th iteration. Then at the $s$-th iteration, we first carrying out the E-step by computing
\begin{eqnarray*}
Q(\bm{\theta}|\bm{\theta}^{(s-1)}) &=& \left. E\Big(\mathcal{L}(\bm{\theta}|\bm{X}, \bm{Y}, \bm{Z})\big|\bm{X}, \bm{Y}\Big) \right|_{\bm{\theta} = \bm{\theta}^{(s-1)}} \\
&=&\sum_{i =1}^n \sum_{\ell =1}^L E\Big(\mbox{I}(D^{(i)} =\ell)|\bm{X}, \bm{Y} \Big) \Big|_{\bm{\theta} = \bm{\theta}^{(s-1)}} \Big(\log(\pi_{\ell})+\log(A_{i, \ell})\Big)\\
&=&\sum_{i =1}^n \sum_{\ell =1}^L \eta_{i, \ell}\Big(\log(\pi_{\ell})+\log(A_{i, \ell})\Big),\\
\end{eqnarray*}
where
\[\eta_{i, \ell} = \frac{A_{i, \ell}\pi_\ell}{\sum_{\ell=1}^L A_{i, \ell}\pi_\ell}\Bigg|_{\bm{\theta} = \bm{\theta}^{(s-1)}}.\]
Then, in the M-step, we compute the maximizer of $Q(\cdot|\bm{\theta}^{(s-1)})$ and take it as the updated estimated parameter vector after the $s$-th iteration. That is,
\[
 \bm{\theta}^{(s)} = \arg\max_{\bm{\theta}} Q(\bm{\theta}|\bm{\theta}^{(s-1)}).
\]

Note that it is not difficult to perform optimization in the M-step. First, we can  apply Lagrange Multipliers method to obtain
\[\pi_\ell^{(s)} = \frac{\sum_{i=1}^n \eta_{i, \ell}}{\sum_{i=1}^n \sum_{j =1}^L \eta_{i, j}}.\]
To find $\bm{\alpha}_{u, v, \ell}^{(s)}$: $u$, $v \in \{ 1, \ldots, K \}$, $\ell \in \{ 1, \ldots, L \}$ so that 
\[
 \sum_{i =1}^n \sum_{\ell =1}^L \eta_{i, \ell} \log(A_{i, \ell})
\]
is maximized, note that
\begin{eqnarray*}
&& \sum_{i =1}^n \sum_{\ell =1}^L \eta_{i, \ell} \log(A_{i, \ell}) \\
&& =  \sum_{\ell =1}^L  \sum_{i =1}^n \eta_{i, \ell} \log(A_{i, \ell}) \\
&& = \sum_{\ell =1}^L \sum_{u=1}^K  \sum_{i =1}^n \sum_{t=2}^T  \sum_{v=1}^K \eta_{i, \ell} \mathbf{I}\Big(Y_{t-1}^{(i)} = u\Big)\mathbf{I}\Big(Y_t^{(i)} = v\Big) \log \left[  P_{v}\Big(\mathbf{X}_t^{(i)}, \bm{\alpha}_{u, 2, \ell}, \ldots, \bm{\alpha}_{u, K, \ell}\Big) \right],
\end{eqnarray*}
so for given $\ell$ and $u$, to find $\bm{\alpha}_{u, 2, \ell}, \ldots, \bm{\alpha}_{u, K, \ell}$ so that
\[
\sum_{i =1}^n \sum_{t=2}^T  \sum_{v=1}^K \eta_{i, \ell} \mathbf{I}\Big(Y_{t-1}^{(i)} = u\Big)\mathbf{I}\Big(Y_t^{(i)} = v\Big) \log \left[  P_{v}\Big(\mathbf{X}_t^{(i)}, \bm{\alpha}_{u, 2, \ell}, \ldots, \bm{\alpha}_{u, K, \ell}\Big) \right]
\]
is maximized is to find the maximum likelihood estimator in a weighted generalized linear model, which can be carried out conveniently using a statistical software such as R.

Once we obtain $\hat{\bm{\theta}}$: the maximum likelihood estimator of the parameter vector in the proposed model, we can perform clustering to the data. For $i \in \{ 1, \ldots, n \}$ and $\ell \in \{ 1, \ldots, L \}$, we can compute the estimated $E\big( \mbox{I}(D^{(i)}=\ell) |\bm{Y} , \bm{X} \big)$: the probability that the $i$-th observation belongs to the $\ell$-th group given the data, which is 
\[  
\frac{A_{i, \ell} \pi_\ell}{\sum_{\ell=1}^L A_{i, \ell}\pi_\ell}\Bigg|_{\bm{\theta} = \hat{\bm{\theta}} }.
\]
We can then perform clustering to the data by assigning the group membership to an observation based on this estimated conditional probability that the observation belongs to a particular group given the data.

\section{Variable selection} \label{sec:var_sel}
Based on our experience of applying the proposed model to analyze some real data set, it is possile to obtain a better description of the data by using the proposed model than using simpler models. However, the number of parameters in the proposed model can be large when there are many explanatory variables. In such case, the parameter estimation can be difficult. To solve this problem, we propose a forward type algorithm for variable selection for the proposed model. We also present the result of a simulation experiment to demonstrate the performance of the variable selection algorithm. We assume that the first explanatory variable is the constant 1 and it is always included in the model.

To describe the proposed forward selection algorithm, we first define some notation.
For two time points $a$ and $b$ such that $1 \leq a<b \leq T$, let $\widehat{\bm{\theta}}_\Lambda^{a:b}$ denote the maximum likelihood estimator of $\bm{\theta}$ based on observations from time $a$ to time $b$ using explanatory variables whose index are in the index set $\Lambda$, and let $\bm{L}_{a:b}$ denote the logarithm of the likelihood function based on observations from time $a$ to time $b$. Then, the proposed forward variable selection algorithm is given below.
\begin{algorithm}[htb] 
\caption{Algorithm for variable selection.} 
\label{alg:Forward} 
\begin{algorithmic}[1]  
\REQUIRE ~~\\  
	Data observed from time 1 to time $T$;\\
	Size of training data $T_1$.
\ENSURE ~~\\  
	The final index set of selected explanatory variables, $\Lambda$.
	\STATE Take the data observed from time 1 to time $T_1$ as the training data and the rest of the data as the testing data.
	\STATE Let $\Lambda_1 = \{1\}$ be the initial index set of selected explanatory variables.
	\STATE For $s = 2, \ldots, p$, let $\Lambda_{s-1}$ denote the index set of selected selected explanatory variables after the $(s-1)$-th iteration.  Let $\Phi = \{1, \ldots, p\}-\Lambda_{s-1}$ and carry out the following steps:
\begin{itemize}
\item Compute
\[
 j^* = \arg \max_{j \in \Phi } \bm{L}_{1:T_1} \left( \widehat{\bm{\theta}}_{\Lambda_{s-1} \cup \{ j \} }^{1:T_1}  \right)
\]
\item Let $\Lambda_s = \Lambda_{s-1} \cup j^*$ and compute
\[
\bm{L}_{(T_1+1):T} \left( \widehat{\bm{\theta}}_{\Lambda_{s}}^{1:T_1}  \right).
\]
\item If \[
\bm{L}_{(T_1+1):T} \left( \widehat{\bm{\theta}}_{\Lambda_{s}}^{1:T_1}  \right) < \bm{L}_{(T_1+1):T} \left( \widehat{\bm{\theta}}_{\Lambda_{s-1}}^{1:T_1}  \right),
\]
then break the for-loop to stop the algorithm and take $\Lambda=\Lambda_{s-1}$.
\end{itemize}
\RETURN $\Lambda$.  
\end{algorithmic}
\end{algorithm}

Next, we will present the result of a simulation experiment to demonstrate the performance of the variable selection algorithm. In the experiment, data were generated from two groups, and each group contained 50 individuals. For each individual, observations for 5  explanatory variables and a response variable at 120 time points were generated. The response variable took values in $\{ 1, 2, 3 \}$. Moreover, the first explanatory variable was the constant 1. The second, 4th and 5th explanatory variables were generated from the standard normal distribution, and the third explanatory variable was generated from the normal distirbution of mean 1 and standard deviation 2. The coefficient vectors $\bm{\alpha}_{u, v, \ell}$s in 
(\ref{mixture1}) used for data generation  are given in Table \ref{table:3}.
\begin{table}[h]
\renewcommand\arraystretch{1.3} 
\begin{center}
\begin{tabular}{cccc}
\hline
\hline
& & $v = 2$ & $v = 3$\\
\multirow{3}{*}{$\ell = 1$} & $u = 1$ & (0.2, 0.2, 1.0, 0.0, 0.0) & (0.4, 0.4, 0.8, 0.0 0.0)\\
& $u = 2$ & (0.1, -0.2, 1.3, 0.0, 0.0) & (0.2, 0.5, 0.5, 0.0, 0.0)\\
& $u = 3$ & (0.2, 0.9, -0.3, 0.0, 0.0) & (0.5, -0.1, 0.3, 0.0, 0.0)\\

\multirow{3}{*}{$\ell = 2$} & $u = 1$ & (0.2, 0.2, 1.3, 0.0, 0.0) & (0.4, 0.3, 0.8, 0.0 0.0)\\
& $u = 2$ & (0.3, 0.2, 1.3, 0.0, 0.0) & (0.2, -0.5, 0.8, 0.0, 0.0)\\
& $u = 3$ & (0.1, -0.5, -0.3, 0.0, 0.0) & (0.5, 0.1, 0.2, 0.0, 0.0)\\
\hline
\hline
\end{tabular}
\end{center}
\caption{the $\bm{\alpha}_{u, v, \ell}$s used in data generation}
\label{table:3}
\end{table}

From Table \ref{table:3}, it is clear that only the first three explanatory variables were important. To examine whether the proposed algorithm is useful for variable selection, we generated 300 data sets using the above data generating process. Given the information that data were from two groups, we were able to choose the important explanatory variables correctly 265 times using the proposed varible selection algorithm, and we chose one more variable 22 times and two more variables 13 times. Therefore, the accuracy rate was over 88\%, which showed that the proposed variable selection algorithm was effective. Thus the variable selection algorithm can be very helpful for model estimation.

\section{Summary}
In this study, a mixture logistic model for panel data with a Markov structure is proposed, and the parameter estimation procedure for the proposed model is also provided. Moreover, to reduce the number of parameters in the model, we propose an algorithm for variable selection. The performance of the variable selection procedure is satisfactory based on our simulation result.

\section*{Acknowledgement}
This research was supported by National Science and Technology Council in Taiwan (MOST 110-2118-M-128-001-).

\end{document}